\documentclass[a4paper,fleqn,usenatbib]{mnras}

\usepackage[T1]{fontenc}
\usepackage{ae,aecompl}

\usepackage{graphicx}	
\usepackage{amsmath}	
\usepackage{amssymb}	
\usepackage{graphics}
\usepackage{epsfig}
\usepackage{txfonts}
\usepackage{flushend}
\usepackage{blindtext}
\usepackage{multirow}

\title[ALMA observations of a metal-rich DLA at $z=2.583$]{ALMA observations of a
metal-rich damped Ly$\alpha$ absorber at $z=2.5832$: evidence for strong
galactic winds in a galaxy group\thanks{This paper makes use
of the following ALMA data: ADS/JAO.ALMA\#2016.1.00628.S (PI Prochaska). ALMA
is a partnership of ESO (representing its member states), NSF (USA) and NINS
(Japan), together with NRC (Canada) and NSC and ASIAA (Taiwan) and KASI
(Republic of Korea), in cooperation with the Republic of Chile. The Joint ALMA
Observatory is operated by ESO, AUI/NRAO and NAOJ. 
Based on observations made with the Nordic Optical Telescope, operated by the 
Nordic Optical Telescope Scientific Association at the Observatorio del Roque 
de los Muchachos, La Palma, Spain, of the Instituto de Astrofisica de Canarias.
Based on observations made with the NASA/ESA Hubble Space Telescope, obtained
at the Space Telescope Science Institute (STScI), which is operated by the
Association of Universities for Research in Astronomy, Inc., under NASA
contract NAS 5-26555. These observations are associated with programme 12553.}}

\author[]{J.~P.~U.~Fynbo,$^{1,2}$ 
K.~E.~Heintz,$^{1,2,3}$
M. Neeleman,$^{4}$
L. Christensen,$^{2}$
\newauthor
M. Dessauges-Zavadsky,$^{5}$
N. Kanekar,$^{6}$
P. M\o ller,$^{7}$
J. X. Prochaska,$^{8}$
N.~H.~P. Rhodin,$^{2}$
\newauthor
M. Zwaan$^{7}$
\\
$^{1}$The Cosmic Dawn Center, Niels Bohr Institute, University of Copenhagen, Juliane Maries Vej 30, 2100 Copenhagen \O, Denmark\\
$^{2}$Dark Cosmology Centre, Niels Bohr Institute, University of Copenhagen, Juliane Maries Vej 30, 2100 Copenhagen \O, Denmark\\
$^{3}$Centre for Astrophysics and Cosmology, Science Institute, University of Iceland, Dunhagi 5, 107 Reykjav\'ik, Iceland\\
$^{4}$Max-Planck-Institut f\"{u}r Astronomie, K\"{o}nigstuhl 17, D-69117, Heidelberg, Germany\\
$^{5}$Observatoire de Gen{\'e}ve, Universit{\'e} de Gen{\'e}ve, 51 Ch. des Maillettes, 1290 Versoix, Switzerland\\
$^{6}$Swarnajayanti Fellow; National Centre for Radio Astrophysics, Tata Institute of Fundamental Research, Pune 411007, India\\
$^{7}$European Southern Observatory, Karl-Schwarzschild Strasse 2, D-85748 Garching, Germany\\
}

\date{Accepted 2018 May 29. Received 2018 May 28; in original form 2018 May 1.}

\pubyear{2018}

\begin{document}
\label{firstpage}
\pagerange{\pageref{firstpage}--\pageref{lastpage}}
\maketitle

\begin{abstract}
We report on the results of a search for CO(3-2) emission from the galaxy counterpart
of a high-metallicity Damped Ly$\alpha$ Absorber (DLA) at $z=2.5832$ towards the quasar Q0918+1636.
We do not detect CO emission from the previously identified DLA
galaxy counterpart. The limit we infer on $M_{gas} / M_{\star}$ is in the low
end of the range found for DLA galaxies, but is still consistent with
what is found for other star-forming galaxies at similar redshifts. Instead we detect
CO(3-2) emission from another intensely star-forming galaxy at
an impact parameter of 117 kpc from the line-of-sight to the quasar and 131 km
s$^{-1}$ redshifted relative to the velocity centroid of the DLA in the quasar
spectrum.  In the velocity profile of the low- and high-ionisation absorption
lines of the DLA there is an absorption component consistent with the redshift
of this CO-emitting galaxy. It is plausible that this component is physically
associated with a strong outflow in the plane of the sky from the CO-emitting
galaxy. If true, this would be further evidence, in addition to what is already
known from studies of Lyman-break galaxies, that galactic outflows can
be traced beyond 100 kpc from star-forming galaxies. The case of this $z=2.583$ 
structure is an illustration of this in a group environment.
\end{abstract}

\begin{keywords}
galaxies: ISM -- ISM: molecules -- quasar: absorption lines -- quasars: individual (Q\,0918+1636) -- submillimeter: ISM
\end{keywords}



\section{Introduction} \label{sec:intro}

The most hydrogen rich class of quasar absorption systems, the DLAs
\citep[defined to have $N$(H\,\textsc{i}) $\geq 2\times 10^{20}$
cm$^{-2}$,][]{Wolfe05}, remain one of the most compelling 
ways to probe the properties of galaxies, in particular at redshifts 2 -- 5
\citep[see][for a review]{Wolfe05}. 
This class of absorbers gives access to detailed information about chemical
evolution. 
The technique is obviously limited to gas-rich galaxies, but is
otherwise not limited to only the brightest galaxies as most other techniques
are \citep{Fynbo08}.

Early on in the studies of DLAs, it was realized that it is important to determine
the emission properties of the DLAs in order to connect the information
collected from absorption studies with the rapidly growing body of information
on high-$z$ galaxies based on emission studies. The Atacama Large Millimeter/sub-millimeter 
Array, ALMA, has opened up a new possibility to do this at sub-millimeter wavelengths. 
An advantage of this approach is that the contrast between the bright background quasar and
the faint DLA galaxy at these wavelengths typically is much smaller than in the optical or 
near-IR bands. The first pilot studies of DLAs with ALMA have given quite interesting results. 
\citet{Moller18} studied a single system at $z=0.716$ and found a surprisingly large
molecular mass and for that mass surprisingly low star-formation rate placing the 
object away from the normal galactic scaling relations for these quantities.
\citet{Kanekar18} studied a sample of DLAs at similar redshifts and found
a surprisingly large detection rate of CO-emission from the galaxy counterparts
of the absorbers. The first reported detection of CO-emission from a
study targetting the field of a $z>2$ DLA is that of \citet{Neeleman18}.
In the field of the $z=2.19289$ DLA towards QSO B1228-113 they detected a 
strong CO-emitter 3.7$''$ (30 kpc) from the quasar sightline.

In this paper we present new observations of the field of the $z=2.5832$ DLA 
absorber towards Q\,0918+1636, which has already been studied in
substantial detail using both imaging and spectroscopy in optical to
near-infrared wavebands. 
The DLA has a high
gas-phase metallicity with measured abundances of [Zn/H] = $-0.12\pm 0.05$ and
[S/H] = $-0.26\pm 0.05$.  The system also shows absorption features from H$_2$
molecules with a column density within the range of $N(\mathrm{H}_2) =
1.5\times 10^{16} - 1.1\times 10^{19}$ cm$^{-2}$ (the large uncertainty
reflecting that the individual absorption components of the H$_2$ lines are not
resolved in the X-shooter spectrum).  The galaxy counterpart is detected
at a projected distance of 1.98 arcsec from the quasar (i.e. 16.2 kpc at
$z=2.5832$). The optical emission lines from [O\,\textsc{ii}],
[O\,\textsc{iii}], H$\beta$ and H$\alpha$ are seen in the combined X-shooter
spectrum, and using emission line diagnostics, \citet[][hereafter F13]{Fynbo13}
found a metallicity of $12 + \log$(O/H) = $8.8 \pm 0.2$ ([X/H]$_{\mathrm{em}}$
= $0.11 \pm 0.20$), consistent with that inferred in absorption (up to the
uncertainties that remain on the absolute calibration of emission line
metallicity diagnostics). The difference between the absorption and emission
redshift amounts to only 36$\pm$20 km s$^{-1}$ (F13). From an Spectral Energy 
Distribution (SED) fit to the optical and near-IR
photometric data, F13 derived a stellar mass of $\log (M_{\star}/M_{\odot}) =
10.10^{+0.17}_{-0.11}$ and a star-formation rate (SFR) of 
$27^{+20}_{-9} M_{\odot}$ yr$^{-1}$ for the DLA emission counterpart. In
addition, the SED fit infers a dust attenuation of $A_V = 1.54^{+0.72}_{-0.56}$
mag. These properties made the system a promising target for our ALMA study of
metal-rich DLAs at $z\gtrsim2$ \citep{Neeleman18}. In this paper we
present ALMA observations of the field targeting the CO(3$-$2) line.

We assume a standard flat cosmology with H$_0$ = 67.8 km s$^{-1}$ Mpc$^{-1}$ , 
$\Omega_m$ = 0.308 and $\Omega_{\Lambda}$ = 0.692 \citep{Planck2016}.

\section{ALMA Observations}

The field surrounding Q0918+1636 was observed with ALMA on UT 2017 January 9
with a compact configuration (maximum baseline of 383 m) for a total on-source 
integration time of 2117 seconds. Quasar J0852+2006 was used for bandpass and 
phase calibration, and quasar J0750+1231 was used for flux calibration. The local
oscillator was tuned so one of the four spectral windows was centered on the 
redshifted CO(3$-$2) line at 96.5 GHz with a frequency resolution of 3.9~MHz. The remaining 
three spectral windows were set up to measure continuum emission of the field. 

The initial data were calibrated using the ALMA pipeline, which is part of the Common
Astronomy Software Applications \citep[CASA;][]{McMullin07} package. After this initial 
round of calibration, additional flagging was performed in CASA. The continuum
image was generated from the three continuum spectral windows, using natural weighting,
resulting in a synthesized beam of $3.''2 \times 2.''5$ at 69$^\circ$. The resulting 
root mean square (RMS) noise of the continuum image is 14.3 $\mu$Jy beam$^{-1}$, no
sources were detected in this image at high signal-to-noise (S/N $> 5\sigma$). 

Using the task \texttt{TCLEAN} in CASA, a spectral cube was made from the spectral 
window centered on the CO(3$-$2) emission at the redshift of the $z=2.5832$ DLA. 
Natural weighting was used and the cube was Hanning-smoothed to a velocity resolution of 
48.5~km~s$^{-1}$, resulting in a RMS noise of 0.24~mJy~beam$^{-1}$ per 48.5~km~s$^{-1}$
channel. The spectral cube spans a velocity window of $\pm 2890$~km~s$^{-1}$ around the
DLA redshift.

\section{Results}

The full ALMA spectral cube was searched for line emission. Only a single line source
was detected at a S/N $> 7$. We do detect one other possible line emission line at $5.8\sigma$
centered at $-700$~km~s$^{-1}$ and $35''$ south-west of the quasar, but no optical counterpart
in the \textit{HST} imaging is seen at this position. Assuming Gaussian noise characteristics, the
chance probability of such a signal to occur in the full data cube due to noise fluctuations is 
$7 \times 10^{-4}$. We note that the signal is detected at $>4\sigma$ in two consecutive, 
independent (24 km~s$^{-1}$) channels, suggesting the emission is real.
However, without a clear optical counterpart, it is hard to interpret this tentative detection.

\subsection{Limit on CO emission from the DLA galaxy counterpart}

At the location of the previously identified DLA galaxy counterpart ---2 arcsec west
of the quasar--- no emission was seen in either the spectral cube or continuum emission.
In the top panel of Fig.~\ref{fig:cospec} we show the ALMA spectrum around the
expected position of the CO(3-2) line.
The $1\sigma$ RMS noise for a 100~km~s$^{-1}$ channel is 0.167 mJy/beam.  Assuming a
similar line width for the emission profile gives a $3\sigma$ upper limit on the velocity-integrated 
line flux of $0.050 \times (\Delta V\,/\,100~\mathrm{km~s}^{-1})^{1/2}~\mathrm{Jy~km~s}^{-1}$. 
At $z=2.5832$ this corresponds to a line luminosity of $L'_{\mathrm{CO}(3-2)} < 1.7\times 
10^9$~K~km~s$^{-1}$ pc$^{2}$, using the conversion formula from \cite{Solomon05}. 
To estimate the molecular mass, we assume $L'_{\mathrm{CO}(3-2)} / L'_{\mathrm{CO}(1-0)} = 0.57$ 
\citep{Dessauges-Zavadsky15}, and a CO-to-H$_2$ conversion of 
$\alpha_{\mathrm{CO}} = 4.3~M_{\odot}$~(K km s$^{-1}$ pc$^2$)$^{-1}$ \citep{Bolatto13},
which results in an upper limit on the molecular gas mass of $M_{\mathrm{mol}} < 
1.3 \times 10^{10} M_{\odot}$. Given the stellar mass of $M_* = 12.6^{+6.1}_{-2.9}\times10^9 M_{\sun}$ 
(from F13) the ratio of $M_{\mathrm{mol}}/M_{\star}$ is still
consistent with the general redshift evolution of gas-to-stellar mass inferred
from main-sequence galaxies \citep[e.g.,][]{Magdis12}. 

For comparison, the $z \sim 0.7$ absorption-selected sample \citep{Moller18,Kanekar18,Klitsch2018} 
have measured molecular masses ranging between $0.6 - 8.2 \times 10^{10} M_\odot$, with 
two non-detections below a molecular mass of $5 \times 10^{9} M_{\odot}$. In addition, the molecular mass
of a $z = 0.101$ galaxy-absorber pair is $M_{\mathrm{mol}} = (4.2\pm 0.2)\times 10^{9} M_{\odot}$ 
\citep{Neeleman16} and the galaxy associated with a $z=2.193$ DLA discussed in \citet{Neeleman18}
has a molecular mass of $(1.4\pm 0.2)\times 10^{11} M_{\odot}$.

\begin{figure*} \centering
\includegraphics[width=2.0\columnwidth]{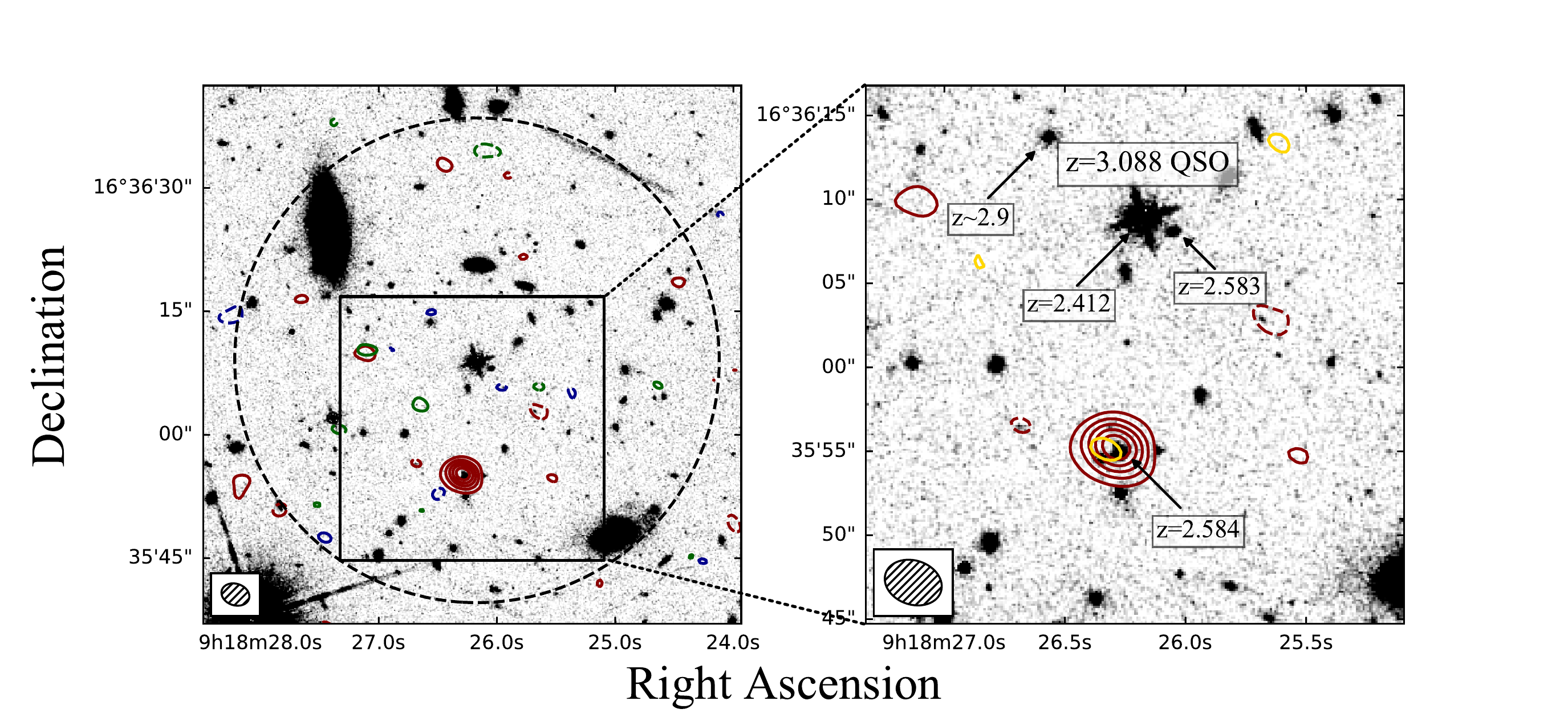}
\vskip 0.3cm
\includegraphics[width=2.0\columnwidth]{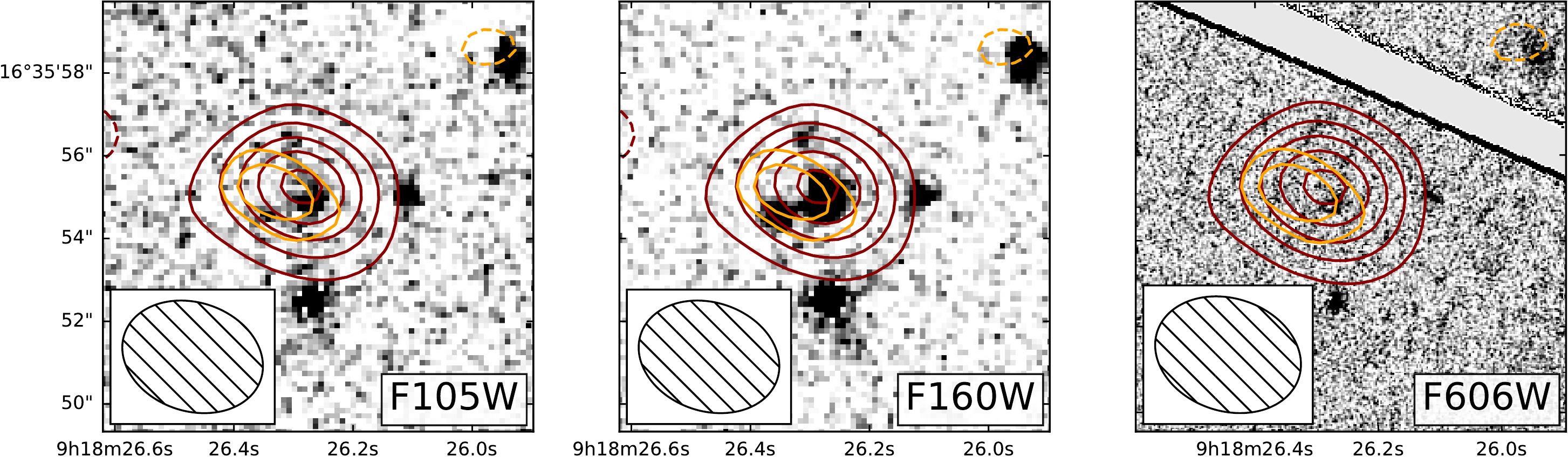}
\caption{\label{fig:ima_hstco} Top panels: Integrated CO(3$-$2) emission (red contours) and
dust continuum (yellow contours) overlaid on the HST/F160W image. The
left panel shows the full field of view of the ALMA observation and the right
panel a zoom on the region around the quasar. No CO emission is detected down
to $0.050 \times (\Delta V\,/\,100~\mathrm{km~s}^{-1})^{1/2}~\mathrm{Jy~km~s}^{-1}$ 
at the position of the previously identified galaxy counterpart 2 arcsec west of the quasar.  
The CO-emitting galaxy is located at a projected distance of 14.2 arcsec (117 kpc at $z=2.5832$)
relative to the quasar.
Bottom panels: 
Integrated CO(3$-$2) emission (red contours) and
dust continuum (orange contours) overlaid on the HST F606W, F105W and F160W
images from F13 from the galaxy located at a distance of 14.2 arcsec (117 kpc
at $z=2.5832$) south of the quasar. The synthesized beam of the ALMA observation is 
shown in the lower left corner of each panel.}
\end{figure*}

\subsection{The CO-emitting galaxy at $\Delta v = 131$ km s$^{-1}$}

\subsubsection{CO emission and molecular gas mass}

While no CO(3$-$2) line emission is detected at the position of the previously identified DLA
counterpart, we do detect a strong CO-emitting galaxy at a projected distance of 14.2 arcsec
(117 physical kpc at $z=2.5832$) south of the DLA (see Fig.~\ref{fig:ima_hstco}). The 
galaxy is also tentatively detected in dust continuum, albeit at a lower signal-to-noise ratio, with a 
measured flux density of 40$\pm$13 $\mu$Jy. The integrated CO(3-2) and dust continuum emission 
from the ALMA observations are shown in Fig.~\ref{fig:ima_hstco}, overlaid on the \textit{HST} images 
of the galaxy from F13. 

In the \textit{HST} F160W image there are multiple emission components near the position of the ALMA 
source. Unfortunately, the resolution of the spectral cube emission is insufficient to resolve the CO 
emission. Together with the known offset in absolute astrometry between ALMA and \textit{HST} 
\citep[e.g.,][]{Dunlop17}, we cannot establish if this is a chance projection of several galaxies
at different redshift or emission from regions in the same galaxy. In the following, we will assume 
the latter. As Fig.~\ref{fig:ima_hstco} shows, the CO emission is co-spatial with a very red component that 
is only seen in the F160W band.

\begin{figure}
	\centering
	\includegraphics[width=\columnwidth]{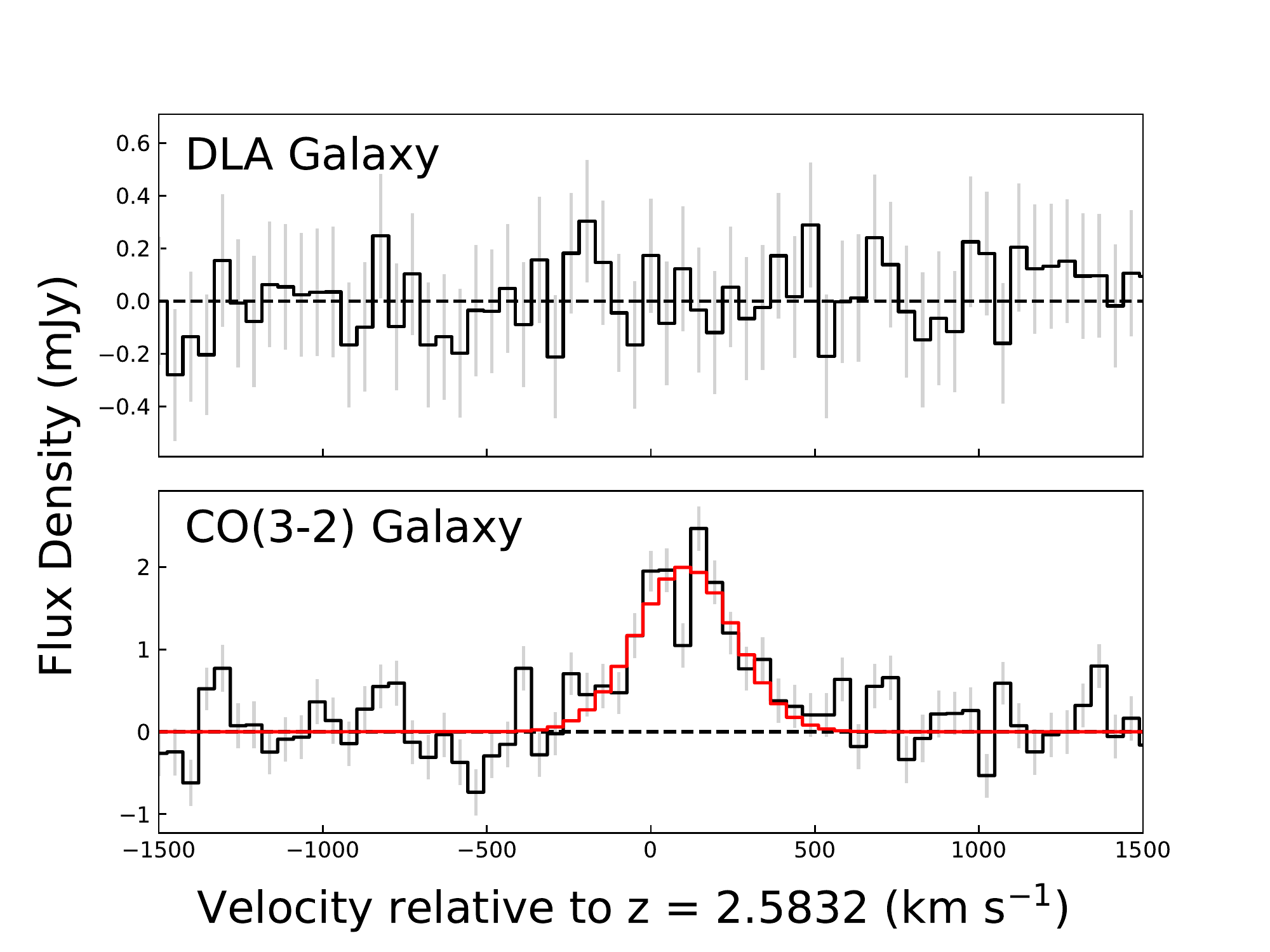}
    \caption{\label{fig:cospec}CO flux density as a function of velocity, where
$v_{\mathrm{rel}} = 0$ km s$^{-1}$ corresponds to $z_{\mathrm{DLA}} = 2.5832$.
The top panel shows the non-detection of the DLA galaxy and the bottom 
panel the spectrum of the CO galaxy 117 kpc from the quasar sightline.
A simple Gaussian profile is fit to the data with an offset of $\delta v = 131$
km s$^{-1}$ relative to $z_{\mathrm{DLA}}$.}
\end{figure}

From the CO spectral cube, we have extracted a line profile at the position of
the CO emission, which is shown in the bottom panel in Fig.~\ref{fig:cospec}. The spectrum shows
that the CO emission is redshifted from the centroid of the DLA at $z=2.5832$
by 131~km~s$^{-1}$. There is marginal evidence for a "boxy" or double-horned
line profile \citep{Davis11}, similar to the CO(2$-$1) and CO(1$-$0) line
profiles observed for two other, low-$z$ absorbing galaxies
\citep{Neeleman16,Moller18}. This indicates that this galaxy has some degree of
a rotational support and/or emission from several sub-clumps.

The velocity-integrated flux density of the CO(3-2) emission line is 
$0.73 \pm 0.08$ Jy km s$^{-1}$. This corresponds to a luminosity of 
$L'_{\mathrm{CO}(3-2)} = (2.5 \pm 0.3) \times 10^{10}$~K~km~s$^{-1}$~pc$^{2}$.


The star-formation rate of this galaxy (see Sect.~\ref{ssec:sed}) is similar to the galaxy
discussed in \citet{Neeleman18}, where we assumed an $\alpha_{\mathrm{CO}}$ = 
4.3$~M_{\odot}$~(K km s$^{-1}$ pc$^2$)$^{-1}$ conversion factor and 
$L'_{\mathrm{CO}(3-2)} / L'_{\mathrm{CO}(1-0)} = 0.57$. To facilitate comparison, we will
use similar values here, although recent work suggest some absorption-selected galaxies 
might show more starburst-like interstellar medium conditions \citep{Klitsch2018}. To account
for this uncertainty, the lower uncertainty on the mass includes the assumption of starburst-like
conditions. Using these conversion factors then yields a molecular gas mass of 
$M_{\mathrm{mol}} = (1.8^{+0.2}_{-1.6})\times 10^{11} M_{\odot}$.

\subsubsection{Associated absorption at the CO galaxy redshift in the quasar spectrum}

\begin{figure}
	\centering
	\includegraphics[width=\columnwidth]{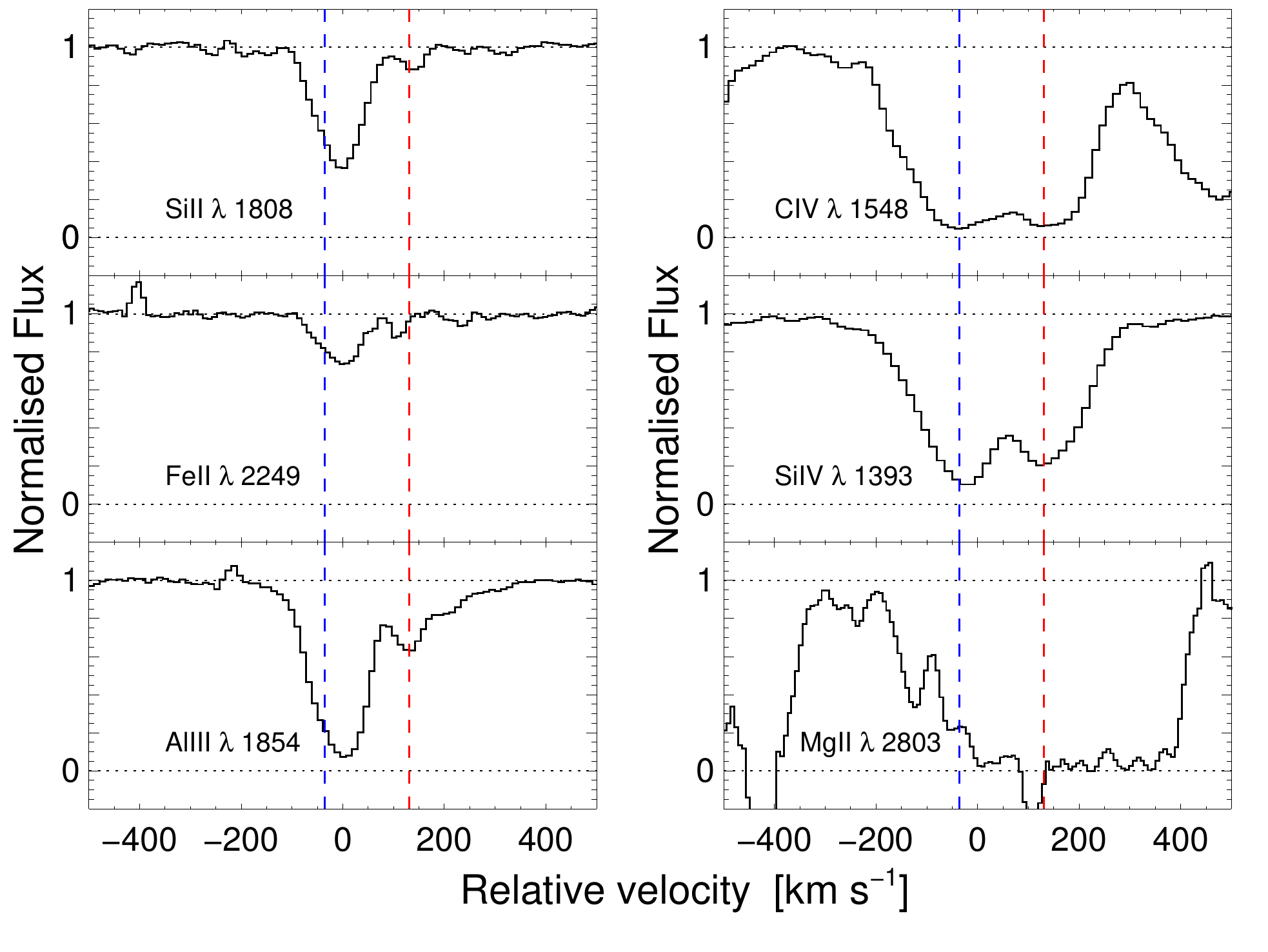}
	\caption{Sections from the normalized X-shooter spectrum of Q\,0918+1636 
     around selected low- and high-ionisation absorption lines from the DLA. 
     The zero-point of the velocity is set to $z_{\mathrm{DLA}} = 2.5832$. 
     The blue dashed line marks the redshift of the previously identified 
     galaxy counterpart at an impact parameter of 16.2 kpc. The red dashed 
     line marks the velocity offset
     of 131 km s$^{-1}$ measured from the centroid of the CO(3-2) emission line
     from the galaxy 117 kpc from the quasar. 
     \label{fig:absline}}
\end{figure}

In Fig.~\ref{fig:absline} we show the normalized quasar spectrum in regions
around selected low- and high-ionisation metal lines from the $z=2.5832$ DLA.
A weak absorption feature is visible at $\Delta v = 131$ km s$^{-1}$ in the 
low-ionisation lines and in the high-ionisation lines there is quite strong 
absorption. 

C\,\textsc{iv} and Si\,\textsc{iv} absorption indicates
> the presence of a warm-hot plasma in galaxy halos or in the IGM and is observed
in most 
quasar and GRB-DLAs \citep[e.g.][and Heintz et al., submitted]{Fox08,Fox09}. 
The large extent of the C\,\textsc{iv} and
Si\,\textsc{iv} absorption line profiles are therefore expected since this gas
traces a more extended medium than the galaxy ISM. In Fig.~\ref{fig:absline} we
also look for extended Mg\,\textsc{ii} absorption due the large stellar mass of
the CO-emitting galaxy (see Sect.~\ref{ssec:sed} below) since the extent of the
Mg\,\textsc{ii} absorbing gas is found to scale with stellar mass and specific
star-formation rate \citep{Chen10}. We do also see extended absorption
in the region of Mg\,\textsc{ii}\,$\lambda\lambda$\,2796, 2803 (albeit in 
a spectral region in the near-IR affected by strong emission lines from 
airglow). There is also
indications of absorption from N\,\textsc{v} and O\,\textsc{vi} in the
spectrum, but both features are heavily blended. It is intriguing that the
high-ionisation metal lines might originate in the IGM between a group of
galaxies at $z=2.5832$. 

We also note that the H$_2$ absorption in the quasar spectrum is distributed over
about 55 km $^{-1}$ in velocity space centred around the mean peak of low-ionisation
absorption \citep[see the lower panel in Fig.~4 in][]{Fynbo11}. Hence, the 
H$_2$ absorption is most likely associated with gas originating in the 
DLA galaxy counterpart 16 kpc from the quasar.

\subsubsection{Physical properties from SED fitting} \label{ssec:sed}

The field of Q\,0918+1636 has been observed with the Wide Field Camera 3 (WFC3)
on-board \textit{HST} in the $F606W$, $F105W$, and $F160W$ filters, and with the 
Andalucia Faint Object Spectrograph and
Camera (ALFOSC) and the Nordic Optical Telescope near-infrared Camera  and
spectrograph (NOTCam) at the NOT (see F13 for details).
From the NOT
we obtained $u$ and $g$ SDSS filter images using ALFOSC and a $K_s$ band image
using NOTCam. 
Using these imaging data we measure the magnitudes in circular apertures with 
a diameter of 2 arcsec for the CO galaxy and
report them in Table~\ref{tab:mag}. For comparison we also list the magnitudes
for the DLA emission counterpart (from F13) and for a galaxy that has been
photometrically determined to be possibly located at a similar redshift (see
Sect.~\ref{ssec:grpgal} below). This galaxy is not detected in the ALMA data.

\begin{table}
\caption{Photometric data for the three galaxies in the field at $z\approx
2.5$ (using a 2 arcsec diameter circular aperture). Magnitudes for the DLA
galaxy are from F13. All magnitudes are given in the AB system and are not
corrected for the Galactic reddening of $E(B-V) = 0.022$ mag
\citep{Schlafly11}.}
	\label{tab:mag}
	\centering
	\begin{tabular}{lccc}
		\noalign{\smallskip} \hline \noalign{\smallskip}
		Band & \multicolumn{3}{ c| }{Source} \\ \noalign{\smallskip}\cline{2-4}\noalign{\smallskip}
		(Mag$_{\mathrm{AB}}$) & DLA galaxy & CO galaxy & Gal. at $z_{\mathrm{phot}} = 2.9^{+0.4}_{-0.9}$  \\
		\noalign{\smallskip}  \hline \noalign{\smallskip} 
		$F606W$ & $25.46\pm 0.13$ & $26.59 \pm 0.25$ & $27.56\pm 0.19$ \\
		$F105W$ & $24.62\pm 0.09$ & $24.70 \pm 0.13$ & $25.34\pm 0.15$ \\
		$F160W$ & $23.68\pm 0.06$ & $23.63 \pm 0.07$ & $23.48\pm 0.07$ \\
		$u$ & $>26.5$ ($3\sigma$) & $>26.5$ ($3\sigma$) & $>26.5$ ($3\sigma$) \\
		$g$ & $25.9\pm 0.3$ & $>26.2$ ($3\sigma$) & $>26.2$ ($3\sigma$) \\
		$K_s$ & $>23.3$ ($3\sigma$) & $>23.3$ ($3\sigma$) & $>23.3$ ($3\sigma$) \\
		\noalign{\smallskip} \hline \noalign{\smallskip}
	\end{tabular}
\end{table}

\begin{figure}
	\centering
	\includegraphics[width=\columnwidth]{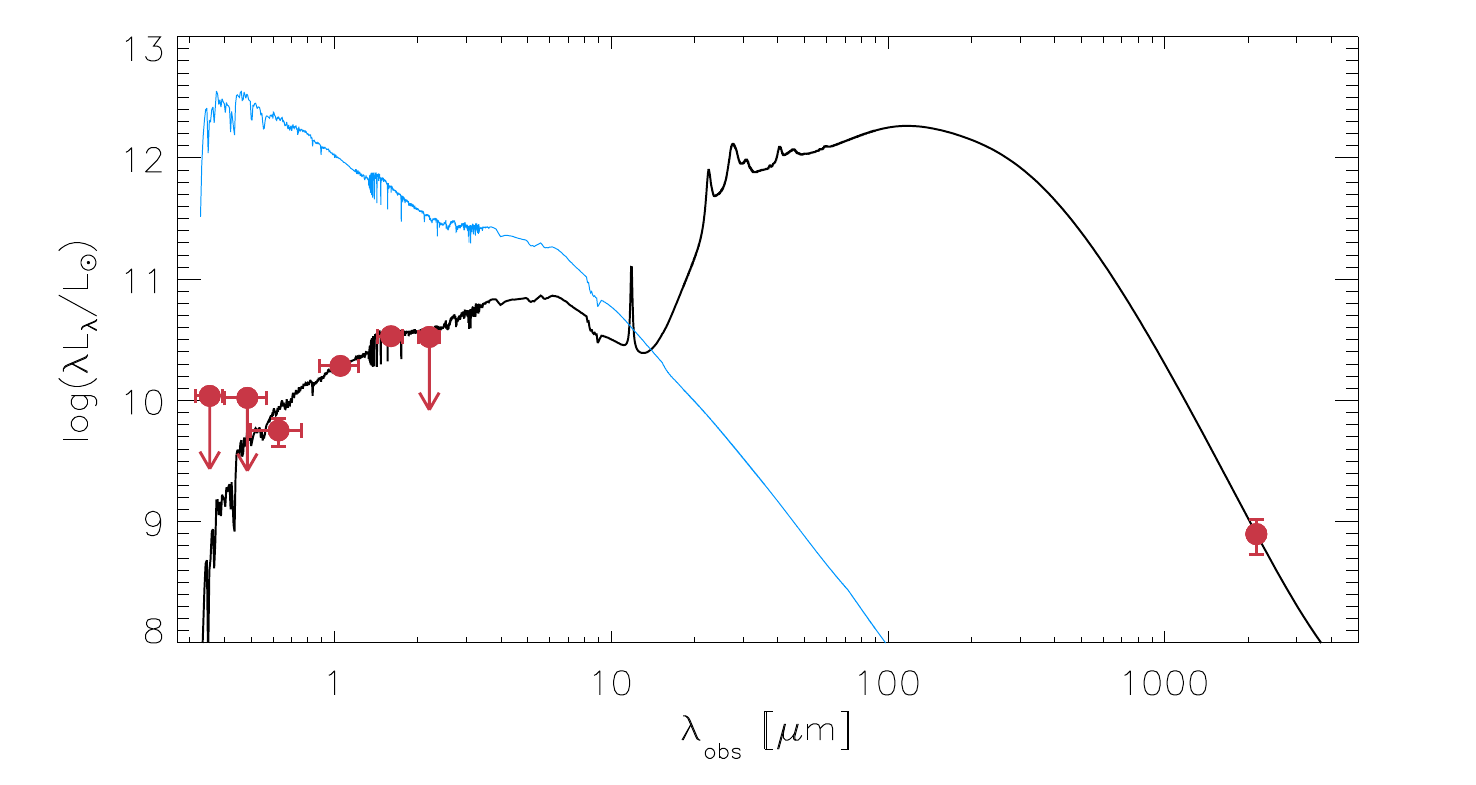}
    \caption{The broad-band optical to near-infrared SED of the CO-emitting
galaxy. The red points denotes the measured photometric data points and upper
limits (arrows). From blue to red wavelengths: ALFOSC $u$ and $g$ bands,
\textit{HST}/WFC3 $F606W$, $F105W$ and $F160W$ bands, the NOTCam $K_s$ band,
and the ALMA continuum detection.
The best-fit galaxy model is shown as the black line. The stellar component
without dust-extinction is shown with a blue line. \label{fig:sedfit}}
\end{figure}

To determine the physical properties of the CO-emitting galaxy we use
\textsc{MagPhys}\footnote{\url{http://www.iap.fr/magphys/}} \citep{daCunha08},
with the photometry in Table~\ref{tab:mag} but corrected for the Galactic
foreground extinction of $E(B-V) = 0.022$ mag \citep{Schlafly11}.
\textsc{MagPhys} is a tool that fits the photometric data to stellar population
and dust emission synthesis models, assuming a \cite{Chabrier03} IMF to
generate the output galaxy models. We also include the detection of the
continuum flux determined from the ALMA data.  The results of the spectral energy
distribution (SED) fits are provided in Table~\ref{tab:sedfit} and the generated galaxy model is illustrated in
Fig.~\ref{fig:sedfit}. The age, extinction and stellar mass of the CO-emitting
galaxy are all similar with those inferred for the DLA galaxy. The SFR,
however, is almost an
order of magnitude higher than that of the DLA galaxy.  
We stress that this photometric modeling is only the best possible with the 
data in hand. Looking at the morphology of the object it clearly consists of
multiple components with different colours so likely both age, the SFR and
the dust vary across the object.



\begin{table}
\caption{Physical properties of the CO galaxy from SED fitting and from
the ALMA observations. 
}
\label{tab:sedfit}
\centering
\begin{tabular}{lccccr}
	\noalign{\smallskip} \hline \noalign{\smallskip}
	Parameter & &    &&& Value \\
	\noalign{\smallskip} \hline \noalign{\smallskip}
	Age (Myr) & &    &&& $217^{+341}_{-137}$ \\
    \noalign{\smallskip}
	$A_V$ (mag) & & &&& $1.88^{+0.50}_{-0.50}$ \\
	\noalign{\smallskip}
	SFR ($M_{\odot}$ yr$^{-1}$) & &&& & $112^{+111}_{-68}$ \\
	\noalign{\smallskip}
	$\log (M_{\star}/M_{\odot}) $ & & &&& $10.45^{+0.10}_{-0.20}$ \\
	\noalign{\smallskip}
	$\log (L_{\mathrm{dust}}/L_{\odot})$ & & &&& $12.05^{+0.30}_{-0.30}$ \\
	\noalign{\smallskip}
	$\log (M_{\mathrm{dust}}/M_{\odot}) $ & & &&& $7.95^{+0.20}_{-0.30}$ \\
	\noalign{\smallskip}
	$\log (M_{\mathrm{mol}}/M_{\odot}) $ & & &&& $11.26^{+0.04}_{-0.95}$ \\
\noalign{\smallskip} \hline \noalign{\smallskip}
\end{tabular}
\end{table}

\subsection{Galaxies in the field around Q\,0918+1636} \label{ssec:grpgal}

In order to search for other possible members of the $z=2.5832$ structure revealed
by the DLA and the CO galaxy we determine photometric redshift measurements of 
galaxies in the overlapping region of the \emph{HST} and NOT imaging data. 
One additional object (marked with $z\sim2.9$ in the upper left panel of Fig.~\ref{fig:ima_hstco},
has a photometric
redshift consistent with 2.6, but we need spectroscopic measurements to 
establish with certainty if this galaxy is really at the same precise redshift.

\section{Discussion}

\begin{figure} \centering \includegraphics[width=\columnwidth]{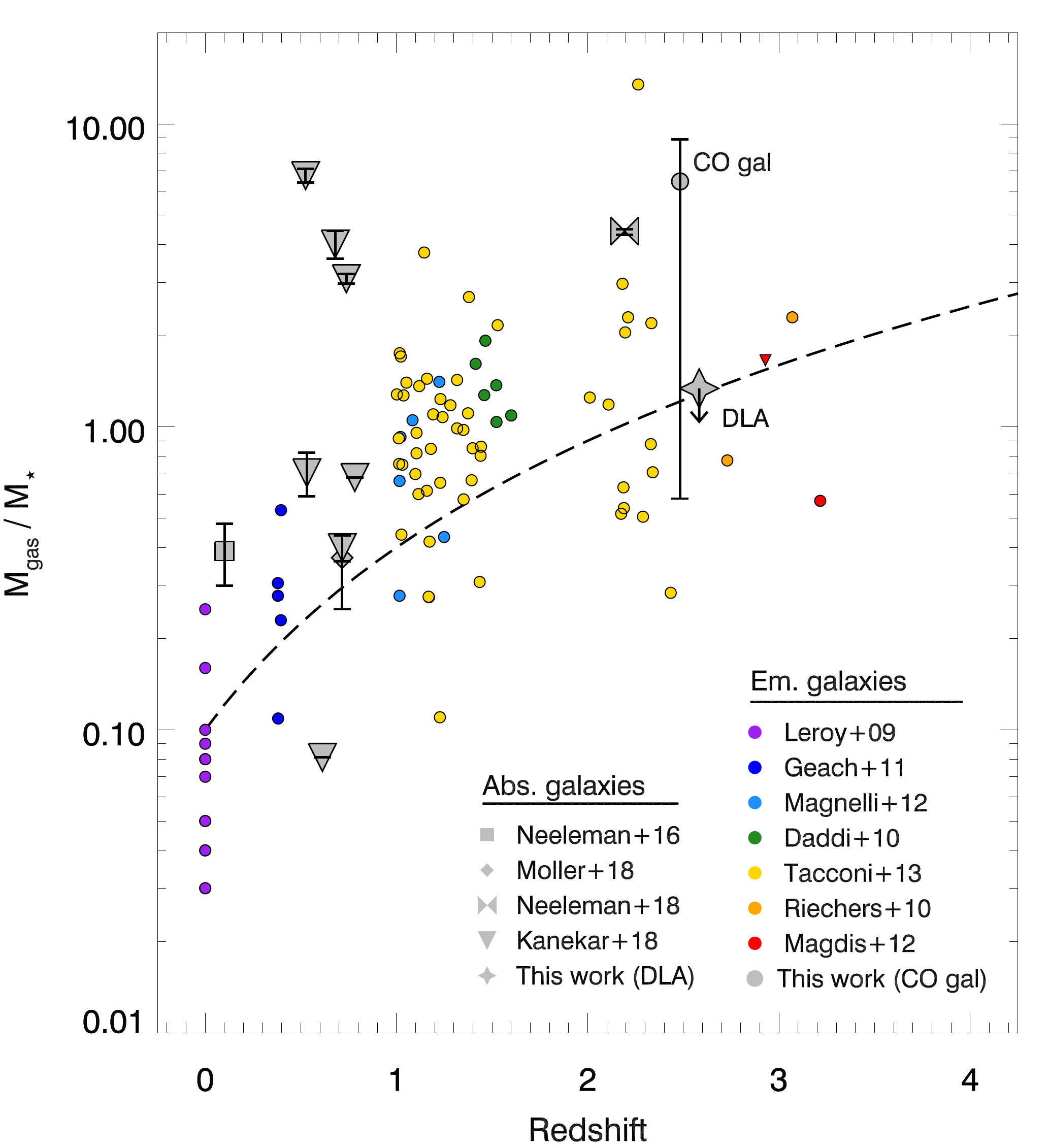}
\caption{The ratio of gas mass to stellar mass as a function of redshift for
various emission (small circles) and absorption selected galaxy samples (gray
symbols). The upper limit for the DLA counterpart and the detetion for the
CO-emitting galaxy in the Q0918+1636 field are the two gray symbols at
$z=2.583$ (the point for the CO-galaxy has been shifted slightly to the left to
increase visibility).  For both the DLA counterpart and the CO-galaxy we show
$M_{gas}/M_{\star}$ assuming $\alpha_{CO} = 4.3$, but the error-bar includes
the possibility for $\alpha_{CO} = 1$. The measurements from other star-forming
galaxies are taken from \citet{Leroy2009,
Geach2011,Magnelli2012,Daddi2010,Tacconi2010,Tacconi2013,Riechers2010}, and
\citet{Magdis2012}. The data for DLA galaxies are from
\citet{Neeleman16,Neeleman18,Moller18}, and
\citet{Kanekar18}.
The dashed curve follows $M_{gas}/M_{\star} = 0.1\times (1+z)^2$ 
\citep[e.g.,][]{Geach2011,CW2013}. \label{fig:fmol}}
\end{figure}

To place the measurements from the field of Q0918+1636 in context we show in Fig.~\ref{fig:fmol} 
our molecular and stellar mass measurements 
in a plot showing the evolution of $M_{gas} / M_{\star}$
as a function of redshift (following \citet{CW2013}). For comparison, we show
measurements from a range of studies of both starformation and absorption selected 
galaxies. The upper limit on the $M_{gas} / M_{\star}$ ratio is in the low end of the 
range found for DLA galaxies, but is otherwise consistent 
with what has been found for star-forming galaxies in general. The non-detection of CO(3-2)
emission is therefore not very surprising for the DLA and it seems that a detection 
will be possibly with a slightly fainter detection limit.

With the detection of the CO-emitting galaxy we can probe this $z=2.5832$
structure at three locations and phases: the metal-rich and H$_2$-bearing DLA
in the quasar spectrum at $z=2.5832$, the previously identified galaxy
counterpart blueshifted by 35 km s$^{-1}$ at an impact parameter of 16.2 kpc
relative to the DLA and now the CO-emitting galaxy redshifted by 131 km
s$^{-1}$ at an impact parameter of 117 physical kpc from the DLA. Photometric
redshifts indicate that there could be other luminous members of the group. There is
evidence from the kinematics that a minor part of the low-ionisation absorption
and a larger fraction of the high-ionisation absorption in the DLA could caused by
gas associated with the CO-emitting galaxy. Whereas we cannot rule out that the 
match is a chance effect, a causal relation seems plausible. The impact parameter of 117 kpc
combined with the age of the star-burst as estimated from SED-fitting requires
a galactic wind velocity of several hundred km s$^{-1}$, which is not
unreasonable \citep[e.g.,][]{Geach2014}.  Together, this shows that the system
is a likely part of a galaxy group and the galactic winds from at least the two
identified galaxies are enriching the group-environment with metals, neutral
hydrogen, dust and molecules \citep[see also][]{JSL2017}.

High-$z$ DLAs have previously been found in environments with other nearby
galaxies
\citep[e.g.,][]{Macchetto93,Moller93,Francis93,Warren96,Moller98,Fynbo03,Schulze12}.
In the work of \citet{Moller93} and \citet{Warren96} a group of three galaxies
were found within 20 arcsec corresponding to about 150 kpc from the quasar
line-of-sight. In this case the DLA is a proximate DLA so the quasar itself
should be included in the structure.  The study of \citet{Fynbo03} found a
large pancake-like structure marked out by 23 Lyman-$\alpha$ emitters in the
field of an intervening DLA at $z=2.85$ towards Q2138$-$4427. There are also
many examples of DLAs or Lyman-limit systems originating from galaxies in group
environments at lower redshifts
\citep{Kacprzak2010,Christensen14,Klitsch2018,Rahmani18}.

We know from a range of different other lines of research that galaxies must expel
large amounts of metals into their environments. In clusters of galaxies the
metals can be directly inferred from observations in the X-ray band of the intracluster
medium \citep[e.g.,][]{Arnaud1992,Alvio1997}. The evidence
supports a scenario, in which the metals in the intracluster medium was 
expelled from 
galaxies at early times \citep[e.g.,][]{Ettori05,Mantz2017}.
The extent of halos of high-ionisation gas has previously been explored 
for Lyman-break galaxies at redshifts of 2.5--3.5
\citep{Adelberger2005}. This study found that strongly star-forming 
galaxies, with typical star-formation rates of several tens of
solar masses per year, are generally associated with haloes of ionised gas traced
by \ion{C}{iv} out to $\sim$80 kpc for $N_{\ion{C}{iv}} \gtrsim 10^{14}$
cm$^{-2}$. This work was extended by \citet{Steidel2010} who studied 
both high- and low-ionisation absorption lines traceable out to impact parameters
of about 100 kpc from $z=2-3$ Lyman-break galaxies.
More recently, \citet{Bielby2017} also explored the correlation 
between Lyman-break galaxies and absorption from neutral hydrogen and
again found a clear correlation with a clustering length of 270$\pm$140 kpc
for \ion{H}{i} absorbers with column densities in the range 
$N_{\ion{H}{i}} = 10^{14.5} - 10^{16.5}$ cm$^{-2}$. Finally we note
that, the DLA studied by \citet{Neeleman2017} also shows evidence for a 
galactic wind given the large impact parameter of the identified galaxy 
counterpart (45 kpc) and a large velocity spreads in the low-ionization metal 
lines.

We know less about the presence of molecules in the circumgalatic medium
at these redshifts. In the present case we know there is H$_2$ at
$z=2.5832$ at several velocity components along the line of sight to the
background quasar. We also now know that there is molecular
gas in the star-burst galaxy 117 kpc from the quasar line of sight.  An
interesting local analogue is M82, from which \citet{Walter2002} detect molecular gas in
the galactic wind more than a kpc away from the disk.  At intermediate
redshifts, \citet{Geach2014} have found a strong molecular outflow from a
starburst galaxy at $z=0.7$ where molecular gas is distributed over more than
10 kpc from an otherwise very compact starburst. \citet{Ginolfi17} discuss an
even more extended molecular gas distribution around a massive star-forming
galaxy at $z=3.47$ extending over more than 40 kpc.  

The fact that both the low- and high-ionisation lines in the quasar spectrum
appear to have contributions from several galaxies, including some at impact
parameters beyond 100 kpc, in a group environment may be part of the reason why
simulations of DLA kinematics have had difficulties matching the large
line-widths of DLAs \citep{X1997,Ledoux1998,Pontzen2008,Barnes2009,Bird2015}. The relatively
strong correlation between metallicity and absorption line widths
\citep{ledoux2006,Moller2013,Neeleman13,Christensen14}, may in addition 
to a mass-metallicity relation, also be partly
influenced by the effect of environment: high metallicity systems will
preferentially trace more biased regions of the Universe with a higher than
average galactic density. 
We finally note that strong feedback, especially for halo masses in the range
10$^{11}$-10$^{12}$ h$^{-1}$ M$_{\sun}$ is required to match 
the column density distribution of DLAs \citep{Bird2014}.  

\subsection{Summary}
In summary, we do not detect CO emission from the previously identified DLA
galaxy counterpart. This non-detection is still consistent with the distribution of 
$M_{gas} / M_{\star}$ found for other star-forming galaxies. 
Instead we detect CO(3-2) from another intensely star-forming galaxy at
an impact parameter of 117 kpc from the line-of-sight to the quasar and 131 km
s$^{-1}$ redshifted relative to the velocity centroid of the DLA in the quasar
spectrum.  In the velocity profile of the low- and high-ionisation absorption
lines of the DLA there is an absorption component consistent with the redshift
of the CO-emitting galaxy. It is plausible that this component is physically
associated with a strong outflow in the plane of the sky from the CO-emitting
galaxy. If true, this would be further evidence, in addition to what is already
known from studies of Lyman-break galaxies, of strong galactic outflows traceable
to impact parameters of at least 100 kpc.

\section*{Acknowledgements}
We thank the anonymous referee for a very constructive and helpful report.
We thank Georgios Magdis for helpful discussions during the preparation of this
manuscript.
The Cosmic Dawn center is funded by the DNRF. KEH acknowledges support by a Project Grant (162948--051) from The Icelandic Research Fund. 
The National Radio Astronomy Observatory is a facility of the National Science
Foundation operated under cooperative agreement by Associated Universities,
Inc. LC and NHR are supported by DFF-4090-00079. MN acknowledges support from
ERC Advanced Grant 740246 (Cosmic\_Gas).




\bibliographystyle{mnras}

\bsp	
\label{lastpage}
\end{document}